%% file: main.tex
\documentclass[sigconf]{acmart}

\usepackage{cleveref}
\usepackage[color=lightgray]{todonotes}
\usepackage[ruled, vlined, linesnumbered]{algorithm2e}
\usepackage{amsfonts}
\usepackage{subcaption}
\usepackage{multirow}
\usepackage{bm}
\usepackage{soul}
\usepackage{bbold}
\usepackage{mathtools}
\usepackage[utf8]{inputenc}
\usepackage{graphicx}
\usepackage{booktabs}
\usepackage{listings}
\usepackage{xcolor}
\usepackage{tabularx}
\usepackage{makecell}
\usepackage{pifont}
\usepackage{siunitx}
\usepackage{csquotes}
\crefname{lstlisting}{algorithm}{algorithms}
\Crefname{lstlisting}{Algorithm}{Algorithms}

\definecolor{codegreen}{rgb}{0,0.6,0}
\definecolor{codegray}{rgb}{0.5,0.5,0.5}
\definecolor{codepurple}{rgb}{0.58,0,0.82}
\definecolor{backcolour}{rgb}{0.95,0.95,0.92}

\lstdefinestyle{mystyle}{
    commentstyle=\color{codegreen},
    keywordstyle=\color{magenta},
    numberstyle=\tiny\color{codegray},
    stringstyle=\color{codepurple},
    basicstyle=\ttfamily\footnotesize,
    breakatwhitespace=false,         
    breaklines=true,                 
    captionpos=b,                    
    keepspaces=true,                 
    showspaces=false,                
    showstringspaces=false,
    showtabs=false,                  
    tabsize=2,
}
\newcommand{\ours}{LlamaRec\xspace}

\AtBeginDocument{%
  \providecommand\BibTeX{{%
    \normalfont B\kern-0.5em{\scshape i\kern-0.25em b}\kern-0.8em\TeX}}}

\setcopyright{acmcopyright}
\copyrightyear{2018}
\acmYear{2018}
\acmDOI{XXXXXXX.XXXXXXX}

\acmConference[Conference acronym 'XX]{Make sure to enter the correct
  conference title from your rights confirmation emai}{June 03--05,
  2018}{Woodstock, NY}
\acmBooktitle{Woodstock '18: ACM Symposium on Neural Gaze Detection,
 June 03--05, 2018, Woodstock, NY} 
\acmPrice{15.00}
\acmISBN{978-1-4503-XXXX-X/18/06}

\begin{document}

\title{\ours: Two-Stage Recommendation using Large Language Models for Ranking}

\author{Zhenrui Yue}
\email{zhenrui3@illinois.edu}
\affiliation{%
  \institution{University of Illinois Urbana-Champaign}
  \city{Champaign}
  \country{USA}
}

\author{Sara Rabhi}
\email{srabhi@nvidia.com}
\affiliation{%
  \institution{NVIDIA}
  \city{Ontario}
  \country{Canada}
}

\author{Gabriel de Souza Pereira Moreira}
\email{gmoreira@nvidia.com}
\affiliation{%
  \institution{NVIDIA}
  \city{São Paulo}
  \country{Brazil}
}

\author{Dong Wang}
\email{dwang24@illinois.edu}
\affiliation{%
  \institution{University of Illinois Urbana-Champaign}
  \city{Champaign}
  \country{USA}
}

\author{Even Oldridge}
\email{eoldridge@nvidia.com}
\affiliation{%
  \institution{NVIDIA}
  \city{British Columbia}
  \country{Canada}
}

\renewcommand{\shortauthors}{Yue, et al.}

\begin{abstract}
Recently, large language models (LLMs) have exhibited significant progress in language understanding and generation. By leveraging textual features, customized LLMs are also applied for recommendation and demonstrate improvements across diverse recommendation scenarios. Yet the majority of existing methods perform training-free recommendation that heavily relies on pretrained knowledge (e.g., movie recommendation). In addition, inference on LLMs is slow due to autoregressive generation, rendering existing methods less effective for real-time recommendation. As such, we propose a two-stage framework using \ul{l}arge \ul{la}nguage \ul{m}odels for r\ul{a}nking-based \ul{rec}ommendation (\ours). In particular, we use small-scale sequential recommenders to retrieve candidates based on the user interaction history. Then, both history and retrieved items are fed to the LLM in text via a carefully designed prompt template. Instead of generating next-item titles, we adopt a verbalizer-based approach that transforms output logits into probability distributions over the candidate items. Therefore, the proposed \ours can efficiently rank items without generating long text. To validate the effectiveness of the proposed framework, we compare against state-of-the-art baseline methods on benchmark datasets. Our experimental results demonstrate the performance of \ours, which consistently achieves superior performance in both recommendation performance and efficiency. 
\end{abstract}

\ccsdesc[500]{Information systems~Language models}
\ccsdesc[500]{Information systems~Recommender systems}
\ccsdesc[500]{Information systems~Personalization}

\keywords{Large Language Models, Recommender Systems}


\received{20 February 2007}
\received[revised]{12 March 2009}
\received[accepted]{5 June 2009}

\maketitle

\input{1_intro}
\input{2_related}
\input{3_method}
\input{4_experiment}

\section{Conclusion}
In this paper, we propose a novel LLM-based two-stage framework \ours for sequential recommendation. The proposed method comprises of two stages: (1)~retrieval stage that adopts sequential recommender to efficiently retrieve candidate items; and (2)~ranking stage, where a LLM-based ranking model is adopted to understand user preference for fine-grained recommendation. Our LLM ranker leverages textual features for preference understanding and is specifically designed to accelerate inference via a simple verbalizer. We demonstrate the effectiveness and efficiency of \ours by performing experiments on benchmark datasets, where \ours consistently achieves superior recommendation results over state-of-the-art baselines. Moreover, \ours exhibits significantly improved inference speed compared to existing generation-based recommenders, showing its potential to further enhance user experience in LLM-based recommendation.

\bibliographystyle{ACM-Reference-Format}
\bibliography{reference}



\end{document}

%% file: 1_intro.tex
\section{Introduction}

Recent advances in large language models (LLMs) have shown significant improvements in various language understanding and generation tasks~\cite{thoppilan2022lamda, chowdhery2022palm, touvron2023llama, OpenAI2023GPT4TR}. By pretraining language models on the next-token prediction task using billions of tokens, LLMs incorporate both extensive knowledge and a wide spectrum of abilities ranging from storytelling to numerical reasoning. For example, the recent Llama~2 model outperforms open-source language models on both human evaluation and benchmark datasets~\cite{touvron2023llama2}. Motivated by such advances, LLMs are also employed in recommendation tasks like retrieval and ranking, showcasing enhanced performance across multiple scenarios~\cite{wang2023zero, chen2023palr, zhang2023recommendation, hou2023large, kang2023llms, liu2023chatgpt, bao2023tallrec, li2023prompt}. 

Nevertheless, the majority of existing works focus on applying LLMs in recommendation and largely ignore the need for efficient inference~\cite{chen2023palr, li2023gpt4rec, wang2023recmind}. In other words, the autoregressive generation of LLMs is often too slow for real-time recommendation, rendering such approaches ineffective for real-world scenarios. To improve recommendation efficiency, one possible solution is to leverage classification or regression heads on top of LLMs to avoid token generation~\cite{kang2023llms}. Yet the heads introduce additional parameters and are trained upon specific settings (e.g., 5-way classification), thereby limiting their applicability for further recommendation tasks. Additionally, many existing studies concentrate on subtasks within recommendation or utilize LLMs for multiple stages, which causes a further reduction in inference speed and thus poses excessive difficulties in achieving efficient recommendation.

In this work, we propose a novel framework for LLM-based two-stage recommendation (\ours), which not only provides a complete solution that includes both retrieval and ranking (also known as recall and (re)rank), but also outperforms existing methods with enhanced recommendation performance and inference efficiency. In particular, we utilize state-of-the-art sequential recommenders to perform ID-based item retrieval, which can efficiently generate candidates regardless of the user history length. Then, we construct the ranking input using a carefully designed template that transforms user history and candidates to text. The constructed text prompt is used to perform parameter-efficient fine-tuning (PEFT) on a pretrained LLM, where the model ranks candidate items by generating scores over candidate indices. Here, we adopt a verbalizer to transform the LLM head output to a probability distribution without additional parameters. The advantages of ranking with our verbalizer approach include: (1)~our LLM achieves significantly reduced inference time by circumventing the autoregressive generation; and (2)~we can generate scores for all candidates within one forward pass and avoid the memory-intensive decoding process (e.g., beam search). As such, \ours can train and infer efficiently in our two-stage framework and provide improved ranking performance compared to state-of-the-art baseline methods.

We summarize our contributions below\footnote{Our implementation is available at https://github.com/Yueeeeeeee/LlamaRec.}:
\begin{enumerate}
    \item We propose a novel framework for LLM-based two-stage recommendation (\ours), which provides a complete solution with both retrieval and ranking.
    \item We use an ID-based sequential recommender as retriever and design a verbalizer approach for LLM-based ranking, which significantly improves the time and memory efficiency for LLM-based recommendation. 
    \item We demonstrate the effectiveness of our \ours on benchmark datasets, where the proposed \ours consistently outperforms baseline methods with considerable improvements in sequential recommendation.
\end{enumerate}

%% file: 2_related.tex
\section{Related Work}

\subsection{Large Language Models}

Recently, substantial progress has been made in the development of large language models (LLMs), with an illustrious example being the launch of ChatGPT\footnote{https://chat.openai.com/}, a powerful LLM-based chatbot. Such advancements in LLMs can be roughly attributed to two main factors: (1)~scaling up the size of language models; and (2)~expanding text corpora in the pretraining stage~\cite{roberts2019exploring, brown2020language, wei2021finetuned, zhang2022opt, thoppilan2022lamda, scao2022bloom, chowdhery2022palm}. Pretrained LLMs leverage self-attention to process input text globally and are optimized via next-token prediction~\cite{vaswani2017attention, radford2018improving}. By incorporating knowledge from the pretraining corpora, LLMs demonstrate advantages in language understanding and generation. Following such paradigm, recent GPT and Llama models~\cite{brown2020language, touvron2023llama, OpenAI2023GPT4TR, touvron2023llama2} show significant improvements on benchmark datasets and human evaluation, and therefore stand as state-of-the-art language models. In this paper, we leverage Llama~2 as our base model and design a efficient tuning and inference framework for two-stage recommendation.

\subsection{LLM-based Recommendation}
LLMs are applied as recommender systems to understand item text features and improve recommendation performance~\cite{geng2022recommendation, li2023text, li2023prompt}. The majority of existing LLM-based recommenders are tuning-free and leverage pretrained knowledge to generate next-item recommendation~\cite{sileo2022zero, wang2023zero, sun2023chatgpt, hou2023large, liu2023chatgpt, wang2023recmind}. For example, Chat-REC~\cite{gao2023chat} utilizes ChatGPT to understand user preferences and improve interactive and explainable recommendation. Another stream of LLM-based recommendation focuses on designing tuning strategies upon subtasks (e.g., rating prediction) to further improve performance~\cite{chen2023palr, kang2023llms, zhang2023recommendation, li2023gpt4rec, petrov2023generative}. For instance, TallRec~\cite{bao2023tallrec} performs instruction-tuning to decide if an item should be recommended. However, most existing works focus on specific recommendation tasks and adopt autoregressive generation to perform inference, leading to substantially increased waiting time. As such, we aim to provide a LLM-based two-stage recommendation framework in \ours, which outperforms existing methods in both performance and efficiency.

%% file: 3_method.tex
\section{Methodology}

\subsection{Setup}
Based on sequential recommendation, our two-stage recommendation framework takes user interaction history $\bm{x}$ from dataset $\mathcal{X}$ as input. In particular, $\bm{x}$ is a sequence of interacted items $[x_1, x_2, \ldots, x_T]$ in chronological order, in which each item is defined in the item space $\mathcal{I}$ ($x_i \in \mathcal{I}, i = 1, 2, \ldots, T$). In the retrieval stage, the items are represented with unique IDs due to large volumes of product data. As for the ranking stage, we leverage item titles to understand user behavior and transition patterns. The output of our framework is the ranking scores $\hat{y} \in \mathbb{R}^{|\mathcal{I}|}$, with the ground truth denoted by $y \in \mathcal{I}$. For optimization, we denote our model with $\bm{f}$ parameterized by $\bm{\theta}$ (i.e., $\hat{y} = \bm{f}(\bm{\theta}; \bm{x})$), which comprises an efficient retrieval model $\bm{f}_{\mathrm{retriever}}$ and a LLM-based ranking model $\bm{f}_{\mathrm{ranker}}$ ($\bm{f} = \bm{f}_{\mathrm{ranker}} \circ \bm{f}_{\mathrm{retriever}}$). Ideally, the highest ranked item in $\hat{y}$ should be the ground truth item $y$ (i.e., $y = \arg\max \hat{y}$). Therefore, the objective of our framework is to maximize ground truth item score. That is, we seek to minimize the expectation of negative log likelihood loss $\mathcal{L}$ w.r.t. parameters $\bm{\theta}$ over $\mathcal{X}$:
\begin{equation}
    \label{eq:transfer_objective}
    \min_{\substack{\bm{\theta}}} \mathbb{E}_{(\bm{x}, y) \sim \mathcal{X}} [\mathcal{L}(\bm{f}(\bm{\theta}; \bm{x}), y)].
\end{equation}

\begin{figure*}[t]
    \centering
    \includegraphics[trim=1cm 2.8cm 1cm 2.8cm, clip, width=0.9\linewidth]{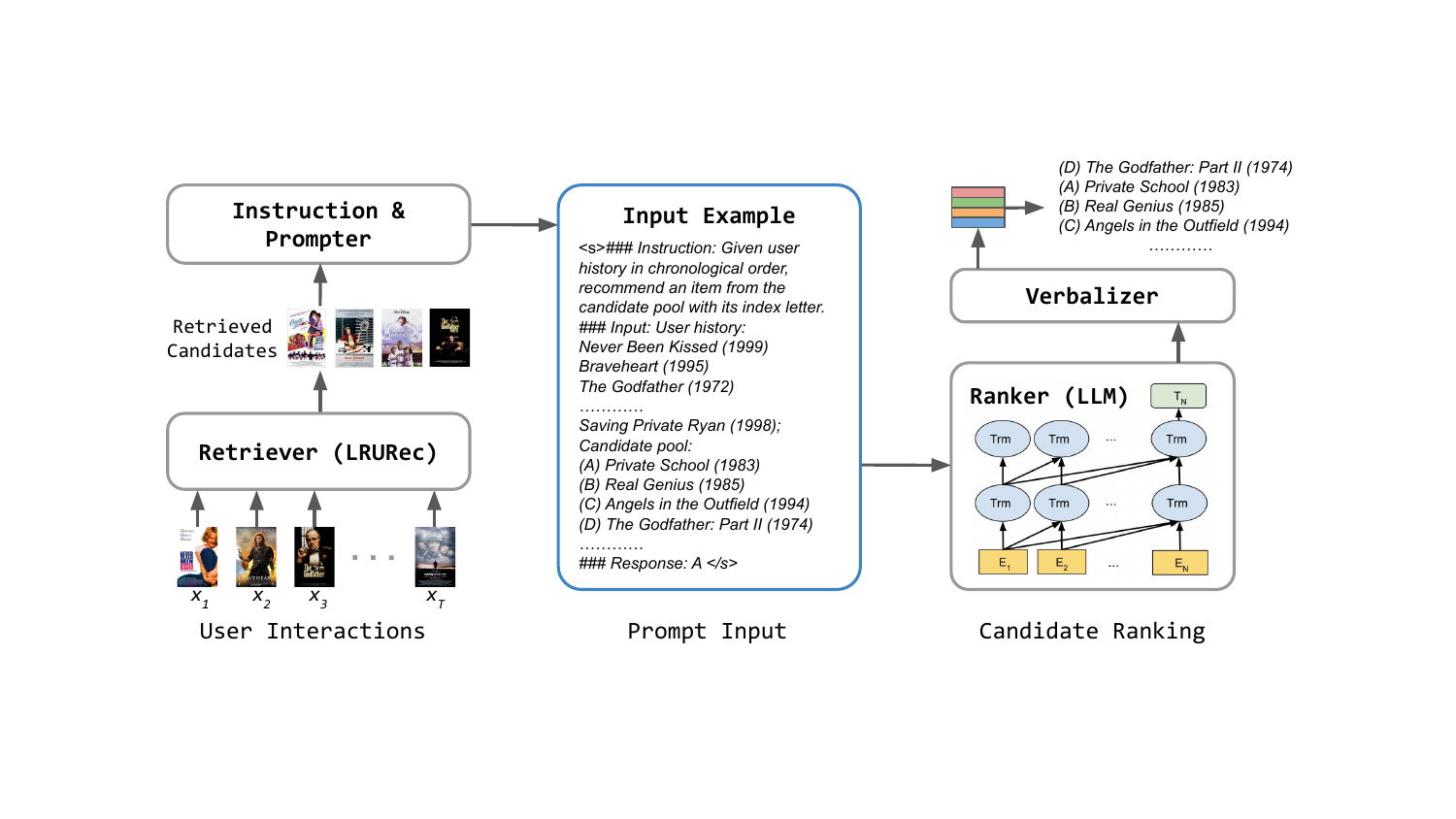}
    \caption{The proposed \ours. The left subfigure illustrates the retrieval stage that generates candidate items with LRURec. Using an instruction template, we transform user history and candidates into text for ranking via Llama~2 (right subfigure).} 
    \label{fig:method}
\end{figure*}

\subsection{The Proposed \ours}
For large-scale recommendation datasets, the size of the item scope $|\mathcal{I}|$ can often be of millions. As such, a common solution is to leverage the two-stage framework, where a fast retriever efficiently generates potential candidates and a more powerful model performs ranking upon the retrieved candidates~\cite{covington2016deep, higley2022building}. Inspired by the two-stage framework, we propose \ours with efficient retrieval and improved ranking via LLM. We illustrate our \ours in \Cref{fig:method} and describe the details of our method in the following.

\subsubsection{Retrieval} 
Since \ours is designed with a two-stage framework, it is possible to select arbitrary model for the retrieval stage. In this work, we adopt the linear recurrence-based LRURec as our retrieval model $\bm{f}_{\mathrm{retriever}}$~\cite{yue2023linear}. LRURec is a small-scale sequential recommender that utilizes linear recurrent units to efficiently process input sequences. LRURec is optimized via autoregressive training to capture user transition patterns and generate predicted item features. For inference, LRURec computes dot product between the item embeddings and the predicted features as item scores. In our \ours, we collect the top-$k$ ($k = 20$ in our experiments) recommendation from LRURec for each input sequence, and the candidate items are saved for the next ranking stage.

\subsubsection{LLM Ranker}
As mentioned, we select Llama~2 as the base model for $\bm{f}_{\mathrm{ranker}}$~\cite{touvron2023llama2}. Specifically, we use the 7B version of Llama~2 to perform ranking among the candidate items from the previous retrieval stage. To construct the text input, we prepend an instruction to describe the task, followed by both history and candidate items represented by their titles. Our prompt template is:
\begin{displayquote}
\#\#\# Instruction: Given user history in chronological order, recommend an item from the candidate pool with its index letter. 

\#\#\# Input: User history: \{ \texttt{history} \}; Candidate pool: \{ \texttt{candidates} \} 

\#\#\# Response: \{ \texttt{label} \}
\end{displayquote}
where \texttt{history}, \texttt{candidates} and \texttt{label} are replaced by history item titles, candidate item titles and the label item of each data example. For inference, the \texttt{label} position is left empty for prediction. We provide an example of our input prompt for movie recommendation in the mid subfigure in \Cref{fig:method}.

\begin{figure}[t]
    \centering
    \includegraphics[trim=4.8cm 5.8cm 4.8cm 5.8cm, clip, width=1.0\linewidth]{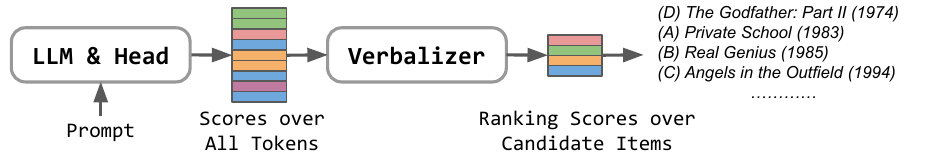}
    \caption{Verbalizer in our \ours ranker.} 
    \label{fig:verbalizer}
\end{figure}

Despite using instructions to prompt LLM, the generated output does not directly provide ranking scores for candidates. To solve this problem, existing works prompt LLMs to generate a ranked list of candidate items~\cite{chen2023palr, hou2023large, liu2023chatgpt, li2023gpt4rec, wang2023recmind}. However, generating a long list is computationally expensive and often requires further processing, as the titles may not exactly match. Unlike such methods, we propose to leverage a simple verbalizer that efficiently transforms the output from the LLM head (i.e., output scores over all tokens) to ranking scores over candidate items (see \Cref{fig:verbalizer}). Specifically, we adopt index letters to identify candidate items (e.g., (A) Private School (1983) (B) Real Genius (1985) (C) ... etc.) and map the ground truth item to the corresponding index letter. Then, the candidate scores can be computed by retrieving the logits of index letters from the LLM head. In other words, the retrieved scores correspond to the next-token probability distribution within the index letters. 

In training, the ground truth item score is maximized by using index letter of the ground truth item as label. As such, our training paradigm is identical to the next-token-prediction task employed by LLMs. Consequently, \ours can be seamlessly combined with arbitrary causal language modeling task to enable multi-tasking capabilities. As for inference, \ours only requires one single forward pass to obtain the LLM head output, followed by retrieving the logits with the proposed verbalizer and ranking the associated items. By using index letters and a simple verbalizer, the LLM model learns to rank items based on user preference while preserving its generative capabilities. Moreover, the ranking inference only needs one forward pass to obtain all scores over candidate items, and thereby significantly improving the ranking efficiency. 

In our \ours implementation, we apply instruction tuning and optimize the model on the response section of the prompt. That is, we only compute loss for \texttt{label} tokens (i.e., index letters and EOS token) in the prompt for each data example. This is because optimizing on the entire input does not yield further improvements, while reducing the loss computation to \texttt{label} section is slightly more efficient in training. To reduce input length of the LLM, we set the maximum value at 20 for user history items and rank the top-20 candidate items from the retriever model. In our experiments, we adopt QLoRA~\cite{dettmers2023qlora} to perform quantization on model parameters for efficient training with reduced computation and carbon footprint. As such, the trainable parameters in our model is less than $1\%$ of the original 7B size and can be performed on consumer GPUs.

%% file: 4_experiment.tex
\section{Experiments}

\begin{table*}[t]
\small
\centering
\makebox[\textwidth]{
\begin{tabular}{@{}lccccc|ccccc|ccccc@{}}
\toprule
\multirow{2}{*}{} & \multicolumn{5}{c|}{\textbf{ML-100k}}                    & \multicolumn{5}{c|}{\textbf{Beauty}}                     & \multicolumn{5}{c}{\textbf{Games}}                       \\ \cmidrule(l){2-16} 
                  & NARM   & BERT   & SAS    & LRU         & \ours           & NARM   & BERT   & SAS    & LRU         & \ours           & NARM   & BERT   & SAS    & LRU         & \ours           \\ \midrule
\textbf{M@5}      & 0.0298 & 0.0188 & 0.0333 & \ul{0.0390} & \textbf{0.0440} & 0.0289 & 0.0246 & 0.0336 & \ul{0.0376} & \textbf{0.0385} & 0.0479 & 0.0422 & 0.0515 & \ul{0.0533} & \textbf{0.0600} \\
\textbf{N@5}      & 0.0359 & 0.0232 & 0.0409 & \ul{0.0468} & \textbf{0.0543} & 0.0342 & 0.0298 & 0.0397 & \ul{0.0435} & \textbf{0.0450} & 0.0576 & 0.0512 & 0.0617 & \ul{0.0640} & \textbf{0.0714} \\
\textbf{R@5}      & 0.0544 & 0.0368 & 0.0641 & \ul{0.0705} & \textbf{0.0852} & 0.0503 & 0.0457 & 0.0582 & \ul{0.0614} & \textbf{0.0648} & 0.0874 & 0.0788 & 0.0930 & \ul{0.0966} & \textbf{0.1061} \\ \midrule
\textbf{M@10}     & 0.0332 & 0.0244 & 0.0378 & \ul{0.0491} & \textbf{0.0529} & 0.0321 & 0.0276 & 0.0371 & \ul{0.0417} & \textbf{0.0428} & 0.0541 & 0.0478 & 0.0583 & \ul{0.0598} & \textbf{0.0671} \\
\textbf{N@10}     & 0.0441 & 0.0366 & 0.0522 & \ul{0.0705} & \textbf{0.0759} & 0.0420 & 0.0372 & 0.0481 & \ul{0.0533} & \textbf{0.0554} & 0.0729 & 0.0649 & 0.0783 & \ul{0.0800} & \textbf{0.0887} \\
\textbf{R@10}     & 0.0801 & 0.0785 & 0.0993 & \ul{0.1426} & \textbf{0.1524} & 0.0746 & 0.0686 & 0.0844 & \ul{0.0916} & \textbf{0.0971} & 0.1351 & 0.1214 & 0.1446 & \ul{0.1463} & \textbf{0.1599} \\ \bottomrule
\end{tabular}
}
\caption{Main recommendation performance, with the best results marked in bold and second best results underlined.}
\label{tab:results}
\end{table*}

\begin{table}[t]
    \small
    \centering
    \begin{tabular}{lrrccc}
    \toprule
    \textbf{Datasets} & \textbf{Users} & \textbf{Items} & \textbf{Interact.} & \textbf{Lenth} & \textbf{Density} \\ \midrule
    \textbf{ML-100k}  & 610            & 3,650          & 100k               & 147.99         & 4e-2             \\
    \textbf{Beauty}   & 22,332         & 12,086         & 198K               & 8.87           & 7e-4             \\
    \textbf{Games}    & 15,264         & 7,676          & 148K               & 9.69           & 1e-3             \\
    \bottomrule
    \end{tabular}
    \caption{Dataset statistics after preprocessing.}
    \label{tab:dataset}
\end{table}

\subsubsection{Datasets}
Our model is evaluated on the following datasets:
\begin{itemize}
    \itemindent=-10pt
    \item \textbf{ML-100k}: A benchmark dataset for movie recommendation with around 100k user-item interactions~\cite{harper2015movielens}.
    \item \textbf{Beauty}: A product review dataset from Amazon website consisting of user feedback on Beauty products~\cite{mcauley2015image, he2016ups}.
    \item \textbf{Games}: A video game dataset from Amazon with user reviews and ratings on video game products~\cite{mcauley2015image, he2016ups}.
\end{itemize}
For preprocessing, we follow~\cite{yue2022defending, chen2023palr} to construct input sequences in chronological order and iteratively filter users and items that are fewer than 5 interactions (i.e., 5-core). Items without meta data (i.e., title) are also filtered. We report the statistics (i.e., users, items, interactions, sequence length and dataset density) in \Cref{tab:dataset}.

\subsubsection{Baseline Methods}
We adopt multiple state-of-the-art sequential recommenders, which include RNN models (i.e., NARM), transformer-based recommenders (i.e., SASRec, BERT4Rec) and linear recurrence-based LRURec. In addition, we adopt LLM-based sequential recommender for comparison in \Cref{sec:llm-baseline}:
\begin{itemize}
    \itemindent=-10pt
    \item \textbf{NARM}: NARM is a RNN-based model that leverages local and global encoder for sequential recommendation~\cite{li2017neural}.
    \item \textbf{SASRec}: SASRec adopts unidirectional attention to process input at a sequence-level to generate next items~\cite{kang2018self}. 
    \item \textbf{BERT4Rec}: A bidirectional attention-based recommender model, BERT4Rec is trained via predicting masked items~\cite{sun2019bert4rec}.
    \item \textbf{LRURec}: An efficient sequential recommender based on linear recurrence, also used as retriever model in \ours~\cite{yue2023linear}.
\end{itemize}

\subsubsection{Evaluation} In our evaluation, we follow the leave-one-out strategy and in each data example, we use the last item for testing, the second last item for validation, and the rest items for training. The evaluation metrics are mean reciprocal rank (MRR@$k$), normalized discounted cumulative gain (NDCG@$k$) and recall (Recall@$k$) with $k \in [5, 10]$. We save the model with best validation scores for evaluation (Recall@10 for retrieval and NDCG@10 for ranking), where predictions are ranked against all items in the dataset.

\subsubsection{Implementation}
For baseline methods and LRURec retriever in \ours, the models are trained with AdamW optimizer using the learning rate of 0.001 and the maximum epoch of 500. Validation is performed every 500 iterations and early stopping is triggered if validation performance does not improve in 20 consecutive rounds. To determine hyperparameters, we perform grid search with weight decay from [0, 1e-2] and dropout rate from [0.1, 0.2, 0.3, 0.4, 0.5]. We used 200 as maximum length for ML-100k and 50 for the other datasets. For our ranker, we use maximum 20 history items and rank the top-20 candidates from the retriever model. The title length is truncated if exceeds 32 tokens. We adopt QLoRA to quantize the Llama~2-based ranker and adopt 8 as LoRA dimension, 32 as $\alpha$ as well as $0.05$ dropout. The LoRA learning rate is 1e-4 with target modules being the $Q$ and $V$ projection matrices. The model is tuned for 1 epoch and validated every 100 iterations. Similarly, the model with the best validation performance is saved for test set evaluation.

\begin{table*}[t]
\small
\centering
\makebox[\textwidth]{
\begin{tabular}{@{}lccccc|ccccc|ccccc@{}}
\toprule
\multirow{2}{*}{} & \multicolumn{5}{c|}{\textbf{ML-100k}}                     & \multicolumn{5}{c|}{\textbf{Beauty}}                      & \multicolumn{5}{c}{\textbf{Games}}                     \\ \cmidrule(l){2-16} 
                  & NARM   & BERT   & SAS    & LRU         & \ours           & NARM   & BERT   & SAS    & LRU         & \ours           & NARM   & BERT   & SAS    & LRU         & \ours           \\ \midrule
\textbf{M@5}      & 0.1369 & 0.0887 & 0.1449 & \ul{0.1965} & \textbf{0.2184} & 0.1961 & 0.1587 & 0.2296 & \ul{0.2944} & \textbf{0.3016} & 0.2039 & 0.1765 & 0.2177 & \ul{0.2504} & \textbf{0.2825} \\
\textbf{N@5}      & 0.1607 & 0.1032 & 0.1793 & \ul{0.2356} & \textbf{0.2693} & 0.2284 & 0.1901 & 0.2679 & \ul{0.3403} & \textbf{0.3524} & 0.2424 & 0.2109 & 0.2571 & \ul{0.3009} & \textbf{0.3360} \\
\textbf{R@5}      & 0.2329 & 0.1477 & 0.2833 & \ul{0.3544} & \textbf{0.4227} & 0.3263 & 0.2861 & 0.3843 & \ul{0.4801} & \textbf{0.5071} & 0.3600 & 0.3160 & 0.3776 & \ul{0.4544} & \textbf{0.4995} \\ \midrule
\textbf{M@10}     & 0.1485 & 0.1054 & 0.1596 & \ul{0.2384} & \textbf{0.2623} & 0.2128 & 0.1743 & 0.2491 & \ul{0.3259} & \textbf{0.3350} & 0.2248 & 0.1947 & 0.2408 & \ul{0.2811} & \textbf{0.3158} \\
\textbf{N@10}     & 0.1886 & 0.1435 & 0.2147 & \ul{0.3367} & \textbf{0.3766} & 0.2689 & 0.2281 & 0.3152 & \ul{0.4168} & \textbf{0.4337} & 0.2931 & 0.2551 & 0.3134 & \ul{0.3760} & \textbf{0.4173} \\
\textbf{R@10}     & 0.3188 & 0.2720 & 0.3927 & \ul{0.6654} & \textbf{0.7560} & 0.4517 & 0.4043 & 0.5312 & \ul{0.7170} & \textbf{0.7600} & 0.5168 & 0.4530 & 0.5521 & \ul{0.6879} & \textbf{0.7522} \\ \bottomrule
\end{tabular}
}
\caption{Recommendation performance on the valid retrieval subset, in which the ground truth item is among the top 20 retrieved items. The best results are marked in bold and second best results are underlined.}
\label{tab:subset-results}
\end{table*}

\subsubsection{Main Results}
We evaluate recommendation performance of \ours and baseline methods, with the results reported in \Cref{tab:results}. Furthermore, we present the performance within the valid retrieval subset. This valid subset only comprises of predictions for which the ground truth item is within the top-20 retrieved items (by $\bm{f}_{\mathrm{retriever}}$), as detailed in \Cref{tab:subset-results}. In both tables, each row represents an evaluation metric and each column stands for one recommender method (and dataset). We use BERT, SAS and LRU to abbreviate BERT4Rec, SASRec and LRURec, while M, N and R stand for MRR, NDCG and Recall respectively. For clarity, we mark the best results in bold and underline the second best results. Notice that the ranking model $\bm{f}_{\mathrm{ranker}}$ only improves recommendation performance within the valid retrieval subset. Based on experiment results, \ours achieves superior performance compared to baseline methods. In particular, we observe: (1)~\ours perform the best across all metrics on all datasets. Compared the the best performing baseline (i.e., LRURec), \ours achieves $11.99\%$, $3.99\%$ and $11.06\%$ average improvements on ML-1M, Beauty and Games respectively. (2)~\ours achieves the largest performance gains on ML-100k, with $12.82\%$, $16.02\%$ and $20.85\%$ improvement on MRR@5, NDCG@5 and Recall@5 respectively. The reason may be attributed to the pretrained movie knowledge and extensive user interactions on ML-100k, where long user history ($\sim$150) allows for a more comprehensive understanding of user preferences. (3)~Within the valid retrieval subset (\Cref{tab:subset-results}), \ours demonstrates larger (absolute) performance gains compared to \Cref{tab:results}. For instance, \ours achieves a further Recall@10 improvement of 0.0643 on Games, compared to only 0.0136 over all user predictions. Overall, the results suggest that \ours can effectively rank candidate items that are of user interest, and thereby improving recommendation performance via language-based preference understanding.

\begin{table}[t]
\small
\centering
\begin{tabular}{@{}lcccccc@{}}
\toprule
\multirow{2}{*}{} & \multicolumn{6}{c}{\textbf{Beauty}}                                               \\ \cmidrule(l){2-7} 
                  & P5     & PALR        & GPT4Rec         & RecMind & POD         & \ours           \\ \midrule
\textbf{N@5}      & 0.0367 & N/A         & N/A             & 0.0289  & \ul{0.0395} & \textbf{0.0450} \\
\textbf{R@5}      & 0.0493 & N/A         & \textbf{0.0653} & 0.0415  & 0.0537      & \ul{0.0648}     \\ \midrule
\textbf{N@10}     & 0.0416 & \ul{0.0446} & N/A             & 0.0375  & 0.0443      & \textbf{0.0554} \\
\textbf{R@10}     & 0.0645 & 0.0721      & \ul{0.0810}     & 0.0574  & 0.0688      & \textbf{0.0971} \\ \bottomrule
\end{tabular}
\caption{Recommendation performance compared to LLM-based baseline methods.}
\label{tab:llm-results}
\end{table}

\subsubsection{Comparison to LLM-based Baselines}
\label{sec:llm-baseline}
We further compare our \ours framework with LLM-based recommendation methods. Among existing works, we find similar studies on LLM-based sequential recommendation, with the majority of them employing Beauty as a standard benchmark. As the implementation is often not available, we compare our results against the reported results from the original works. The adopted LLM-based methods include: P5, PALR, GPT4Rec, RecMind and POD~\cite{geng2022recommendation, chen2023palr, li2023gpt4rec, wang2023recmind, li2023prompt}. The comparison results are reported in \Cref{tab:llm-results} in a similar fashion to \Cref{tab:results}. Surprisingly, we observe significant improvements using \ours on Beauty, with an average gain of $14.31\%$ compared to the second best results across all metrics. For instance, \ours can achieve $24.22\%$ performance improvement on NDCG@10 against the best-performing baseline PALR. In the case of Recall@5, GPT4Rec exhibits a slight performance advantage of 0.0005 over \ours, yet it's important to note that the metrics of GPT4Rec are computed based on five queries against one forward pass in \ours. In summary, the evaluation results against existing LLM-based methods indicate the effectiveness of our two-stage design and verify the effectiveness of \ours in both retrieval and ranking.

\subsubsection{Efficiency of \ours}
\begin{figure}[t]
    \centering
    \includegraphics[trim=1.2cm 0.1cm 3cm 1.1cm, clip, width=0.75\linewidth]{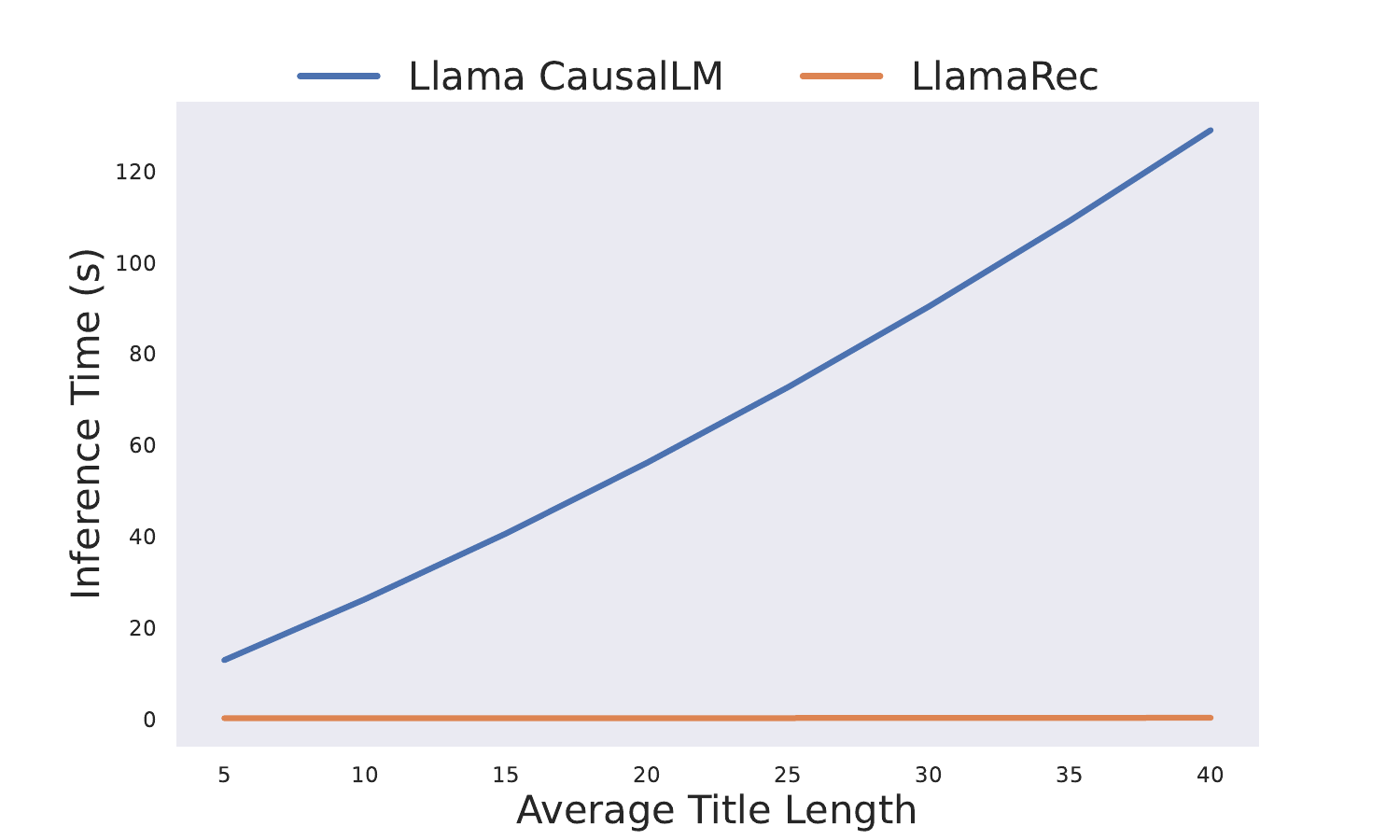}
    \caption{Inference Efficiency of Llama and \ours.} 
    \label{fig:inference}
\end{figure}

We now evaluate the ranking efficiency of our verbalizer approach in comparison to the generation approach. For our baseline, we adopt the same Llama~2 model (Llama CausalLM) and vary candidate title lengths in the prompt to generate a ranked list of the candidate titles (as in~\cite{chen2023palr, hou2023large, liu2023chatgpt, li2023gpt4rec, wang2023recmind}). In response generation, we perform greedy search for decoding and terminate the generation if response length exceeds the length of all titles combined. In contrast to Llama CausalLM, \ours only need one forward pass to obtain the ranking scores over all candidate items. We present the visualized results of inference time in \Cref{fig:inference}, with x-axis and y-axis representing the average title length and time (in s). As expected, we observe significantly improved efficiency using \ours. For example, the inference time of \ours is consistently under 1s regardless of title length, whereas for an average title length of 20, the generation approach takes 56.16s inference time. In sum, the efficiency of \ours outperforms the baseline generation method by a large margin, showing its potential in real-world recommendation scenarios.

